# Mechanisms of photoconductivity in atomically thin MoS$_2$

Marco M. Furchi, Dmitry K. Polyushkin, Andreas Pospischil, and Thomas Mueller[*]

*Vienna University of Technology, Institute of Photonics,*

*Gußhausstraße 27-29, 1040 Vienna, Austria*

**ABSTRACT: Atomically thin transition metal dichalcogenides have emerged as promising candidates for sensitive photodetection. Here, we report a photoconductivity study of biased mono- and bilayer molybdenum disulphide field-effect transistors. We identify photovoltaic and photoconductive effects, which both show strong photogain. The photovoltaic effect is described as a shift in transistor threshold voltage due to charge transfer from the channel to nearby molecules, including SiO$_2$ surface-bound water. The photoconductive effect is attributed to trapping of carriers in band tail states in the molybdenum disulphide itself. A simple model is presented that reproduces our experimental observations, such as the dependence on incident optical power and gate voltage. Our findings offer design and engineering strategies for atomically thin molybdenum disulphide photodetectors and we anticipate that the results are generalizable to other transition metal dichalcogenides as well.**

**KEYWORDS:** *transition-metal dichalcogenides, molybdenum disulphide, photoconductive effect, photovoltaic effect*

The photoresponse of two-dimensional (2D) materials, such as graphene[1] and layered transition metal dichalcogenides[2] (TMDs), are currently the subject of intensive research[3–26]. Although graphene offers the possibility of photodetection over an unrivaled wavelength range[3–5] and with ultra-high bandwidth[5, 6], its photoresponsivity is limited by the vanishing bandgap and picosecond carrier lifetime. Recently, TMDs have attracted much interest for photodetection applications[11–26], due to their potential for achieving very high responsivities. Broadly speaking, two groups of TMD-based photodetectors have been investigated: (i) vertically stacked heterostructure devices, where a few-layer TMD semiconductor is sandwiched between graphene or transparent metal electrodes[11, 12], and (ii) lateral metal-TMD-metal detectors[13–24], whose device structure resembles that of field-effect transistors[27–29] (FETs). The latter, are mostly operated as photoconductors (i.e. with applied bias voltage) which allows overcoming the major drawback of photodiodes: unity gain. In molybdenum disulfide ($MoS_2$) transistors, for example, responsivities in the order of $\sim 10^3$ A/W have been achieved[16, 17]. On the other hand, other work has reported much lower responsivity values of[14] 0.1 A/W or[13] 0.008 A/W only. Also, a strong variation in response time has been obtained, ranging from microseconds[18, 21] to seconds[16, 17]. Such a strong variation in performance despite similar device layouts is rather peculiar, and requires further investigation. By comparison of results obtained from samples with different geometries, device operation under various conditions, and theoretical modeling, we are able to elucidate the physical mechanisms that give rise to photoconductivity and photogain in $MoS_2$ transistors.

In our work, we investigated back-gated FETs[27–29] (Figure 1a) with monolayer and bilayer $MoS_2$ channels. The devices were fabricated by mechanical exfoliation of natural MoS$_2$ (*SPI Supplies*) onto a Si/SiO$_2$ wafer with $d_{ox}$ = 280 nm oxide thickness. Standard photolithographic and metal evaporation techniques were employed to produce the drain and source contact electrodes (5 nm Ti, 50 nm Au). Table S1 in the Supporting Information provides basic device parameters, such as dimensions and mobilities. A microscope image of one of our devices (Sample 7) is shown in Figure 1b. Care was taken to select flakes with approximately rectangular shape, so that channel length *L* and channel width *W* are well defined. Raman spectroscopy was employed to determine the MoS$_2$ thickness via the energy difference between the $E^1_{2g}$ and $A_{1g}$ Raman modes[30, 31] (Figure 1c). MoS$_2$ flakes with energetic $E^1_{2g}$-$A_{1g}$ splittings of ~19 cm$^{-1}$ were identified to be monolayers, those with ~21 cm$^{-1}$ were determined to be bilayers. The Raman measurements were accompanied by optical contrast and, in part, atomic force microscopy (AFM) measurements. Prior to all measurements, the samples were annealed in vacuum (~5×10$^{-6}$ mbar) at 400 K for several hours to remove photoresist residues, water, and other contaminations from the surface.

In a first step after device fabrication, we performed electrical characterization using a semiconductor parameter analyzer (Agilent 4155C) with the samples mounted in a wafer probe station (Lakeshore). The optical viewport of the probe station was covered in order to avoid illumination by ambient light. The electrical measurements – as well as all other measurements presented in this letter – were performed at room temperature (*T* = 300 K). The gate characteristic $I_D$–$V_G$ of Sample 7 is presented in Figure 1d ($V_D$ = 2 V). The curve resembles that of a typical n-type MoS$_2$ FET, in agreement with numerous previous reports[27–29]. Using the expression $\mu_n = (L/W)\, C_{ox}^{-1}\, V_D^{-1}\, (dI_D/dV_G)$, where $C_{ox} = \epsilon_{ox}/d_{ox}$ = 115 µF m$^{-2}$ is the oxide capacitance per unit area and $\epsilon_{ox}$ is the SiO$_2$ permittivity, we extract electron field-effect mobilities between $\mu_n$ = 0.9 and 7.6 cm$^2$/Vs (see Supporting Information, Table S1). From the $I_D$–$V_G$ curve, plotted on a logarithmic scale, we determine a subthreshold swing of $S = dV_G/d(\log_{10} I_D) \sim 2$ V/decade, evidencing the existence of trap states[28] – a common problem in MoS$_2$ transistors. Figures 1e and f show $I_D$–$V_D$ curves for gate voltages between -80 and 80 V. The curves exhibit ohmic characteristics for all gate voltages, indicating the absence of blocking (Schottky) contacts. The influence of the contacts on photoconductivity (due to photo-induced Schottky barrier lowering[32]) can thus be disregarded in our devices. Moreover, no current saturation is observed and we can calculate the drain current from Ohm's law $I_D = (W/L)V_D\sigma$, with $\sigma = q\mu_n n$ being the conductivity, $q$ the electron charge, and $n$ the electron density.

The photoresponse of nanomaterials often occurs due to various mechanisms that act simultaneously, making it difficult to disentangle the individual contributions. Graphene can serve as an example – there, the relative importance of the photoelectric (PE) versus the photo-thermoelectric (PTE) effect is still being discussed[10, 33]. PE[22] and PTE[23] effects have also been reported for MoS$_2$, and although these mechanisms dominate the photoresponse at zero bias, they both can be neglected under the operational conditions presented in this Letter (see Supporting Information, Note S1). In general, two mechanisms can give rise to photoconductivity in a transistor: the *photovoltaic* (PV) and the *photoconductive* (PC) effects[34–36]. The total photocurrent $I_{ph}$, defined as difference between device current under illumination and dark current, $I_{ph} = I_{D,illum} - I_{D,dark}$, is the sum of these two contributions,

$$I_{ph} = I_{ph,PV} + I_{ph,PC}. \quad (1)$$

The PV effect can be described as the change in transistor threshold voltage $V_T$ (in the dark) to $V_T - \Delta V_T$ (under optical illumination) and is therefore sometimes referred to as "photogating" effect. As a result of threshold shift, the drain current changes from $I_D$ to $I_D + \Delta I_D$ and it follows that[33] $I_{ph,PV} = I_D(V_G - V_T + \Delta V_T) - I_D(V_G - V_T)$, or

$$I_{ph,PV} \approx g_m \Delta V_T, \quad (2)$$

with $g_m = dI_D/dV_G$ being the transconductance. The PC component of the photoresponse is the increase in conductivity $\Delta \sigma$ due to photoinduced excess carriers,

$$I_{ph,PC} = \left(\frac{W}{L}\right) V_D \Delta \sigma. \quad (3)$$

We identify both effects in our devices and will now present experimental results and a quantitative analysis.

**Photovoltaic Effect**

Figure 2a shows $I_D$–$V_G$ curves ($V_D$ = 5 V) of Sample 4 in the dark (dash-dotted blue line) and under illumination with white-light from a halogen lamp (dashed red line). The data were acquired by purging the wafer probe station with nitrogen gas ($N_2$). The photocurrent $I_{ph}$ is obtained by subtracting one curve from the other and is depicted as solid black line. $I_{ph}$ = 3.1 µA at $V_G$ = 100 V. Given the low power density of the optical illumination (the total power impingent on the device is $P_{opt} \sim 5$ nW), this value translates into a remarkably high photoresponsivity of $R = I_{ph}/P_{opt} \sim 10^3$ A/W, in agreement with results recently reported for CVD-grown[17] and exfoliated[16] monolayer MoS$_2$ FETs.

In the same graph we plot the transconductance $g_m$ on the right axis (symbols). The similarity between $I_{ph}$ and $g_m$ is striking and we conclude that the photoresponse is dominated by the PV effect, because the PC effect is not expected to show an appreciable dependence on gate voltage. Using equation (2), we extract $\Delta V_T$ = 22 V. Similar results were obtained in vacuum ($\Delta V_T$ = 18 V) and, unless otherwise stated, all further measurements were thus performed in $N_2$-atmosphere.

In the inset of Figure 2a we present the power dependence of $\Delta V_T$ (measured with a solid-state laser at $\lambda$ = 532 nm wavelength). It rises steeply for optical powers up to ~1 nW and then saturates. The dynamics of the photocurrent, plotted in the inset of Figure 2b, shows a complex multi-exponential rise/decay behavior. We determine the dominant time constant $\tau_S$ ~ 2 s from the time after which the signal reaches 63 % of its maximum value. From the absorbed photon flux $\phi = \eta P_{opt}\lambda/(Ahc)$, where $\eta$ ~ 0.15 is the optical absorption of $MoS_2$ in the visible[2], $A$ is the area of illumination, $h$ is Planck's constant, and $c$ denotes the speed of light, the trapped charge density at saturation can roughly be estimated as $N_{sat} = \phi_{sat}\tau_S$ ~ 7.5×10$^{15}$ cm$^{-2}$. The thermally grown $SiO_2$ surface consists of Si-OH silanol groups with a density of ~10$^{15}$ cm$^{-2}$ that are hydrated by water molecules[37]. It is thus likely that the PV effect arises from charge trapping by few layers of surface-bound water underneath the $MoS_2$ sheet. Upon changing the environment from $N_2$ to air (Figure 2b), we observe a noticeable increase of $\Delta V_T$ to ~150 V due to additional surface water and other adsorbents.

**Photoconductive Effect**

Next, $I_D$-time traces, illustrated in the inset in Figure 3a, were taken by illuminating the device with a chopped laser beam ($\lambda$ = 532 nm; $P_{opt}$ = 1 μW; modulation frequency $f$ = 3 kHz) and recording the drain current with an oscilloscope (dark current: red dashed line;

with optical illumination: blue solid line). We observe a strong background photocurrent, and, superimposed, a much smaller current that oscillates with the modulation frequency. After blocking the light, the oscillating component disappears immediately and the background signal drifts slowly (on a time scale of seconds) back to the initial value. The slow signal is the contribution due to the PV effect described above. We argue that the fast photoresponse is of different physical origin because of its weak gate voltage dependence (Figure 3e). We also have not observed any substantial dependence of the photocurrent amplitude on the gaseous environment and modulation frequency (Figure 3a).

We attribute the fast component to the PC effect and will now substantiate this claim. All further measurements were performed using a lock-in detection scheme. This technique is sensitive to the oscillating signal only, allowing us to effectively suppress the strong (and slow) PV response. In our setup light from a solid-state laser ($\lambda = 532$ nm) was focused to a ~0.8 μm (full-width at half-maximum) diameter spot using a microscope objective. A piezo-electrically driven tilt mirror, mounted before the objective, allowed us to control the beam position on the sample. The incident optical power on the sample was varied using a variable attenuator and neutral density filters and the electrical output signal was fed into a low-noise current pre-amplifier (Stanford Research, SR570). The amplifier also allowed us to apply the drain voltage $V_D$ necessary for photoconductivity measurements. Its output signal was detected with the lock-in and the laser beam was chopped with a mechanical chopper at frequencies between 1 and 3 kHz. Photocurrent maps ($\lambda = 532$ nm) are presented in Supporting Information, Figure S1. Measurements performed with photon energies below the $MoS_2$ bandgap ($\lambda = 830$ nm) resulted in orders of magnitude lower signal levels, confirming that the response stems from light absorption in MoS$_2$. Figures 3d and e show the photocurrent amplitudes (extracted from the photocurrent images) as a function of incident optical power and gate voltage (Sample 7).

It is instructive to consider the case of a trap-free semiconductor first. From the photon flux $\phi$, we estimate the density of photogenerated electrons $\Delta n$ and holes $\Delta p$ by $\Delta n = \Delta p = \phi \tau_r$, where $\tau_r \sim 100$ ps is the (non-radiative) carrier lifetime[38]. The photocurrent is calculated from Equation (3) with $\Delta \sigma = q(\mu_n + \mu_p)\Delta p$, where $\mu_p = \mu_n$ is assumed. For the responsivity we then obtain $R \sim 0.06$ A/W, which is two orders of magnitude smaller than the experimentally observed value in Figure 3d (~6 A/W at low incident optical power).

It is thus evident that the device exhibits photoconductive gain, which is typically due to the trapping of one type of charge carrier. Previous work[39–41] has confirmed the presence of band tail states in the MoS$_2$ conduction and valence bands as a result of structural defects in MoS$_2$ itself or induced by disorder[40]. In addition to these shallow traps, deep recombination centers (mid-gap states) are existent[41] which give rise to non-radiative (Shockley-Read-Hall-type) recombination with rate $\tau_r^{-1}$. In Figure 3b we schematically illustrate the density of states in MoS$_2$. In order to calculate the photoconductive response in the presence of traps, we adapt a modified Hornbeck-Haynes model[42–44] (Figure 3c) in which we approximate the valence band tail by a narrow distribution of states with density $P_t$ and with energy $E_{V,t}$ above the valence band edge $E_V$. Trapping and escape of holes into/from these states occurs with rates $\tau_t^{-1}$ and $\tau_g^{-1}$. For the moment, we neglect electron trapping, as we perform our analysis in the transistor ON-state ($V_G > V_T$), in which the Fermi level $E_F$ resides in the vicinity of the conduction band edge $E_C$ so that most of the electron traps are filled.

Under optical illumination, charge neutrality demands that $\Delta n = \Delta p + p_t$, where $p_t$ is the trapped hole carrier concentration and $\Delta n$ and $\Delta p$ denote the free carrier concentrations. For the excess conductivity we thus obtain

$$\Delta \sigma = q(\mu_n + \mu_p)\Delta p + q\mu_n p_t, \qquad (4)$$

which is by the amount $q\mu_n p_t$ higher than what would be expected without traps. The differential equations describing the carrier dynamics,

$$\frac{d\Delta p}{dt} = \phi - \frac{\Delta p}{\tau_r} + \frac{p_t}{\tau_g} - \frac{\Delta p}{\tau_t}\left(1 - \frac{p_t}{P_t}\right) \qquad (5)$$

$$\frac{dp_t}{dt} = -\frac{p_t}{\tau_g} + \frac{\Delta p}{\tau_t}\left(1 - \frac{p_t}{P_t}\right) \qquad (6)$$

can readily be solved under steady-state conditions, $d\Delta p/dt = dp_t/dt = 0$, to obtain

$$\Delta p = \phi \tau_r, \qquad (7)$$

$$p_t = \frac{\phi P_t \tau_r}{\phi \tau_r + P_t\left(\frac{\tau_t}{\tau_g}\right)}. \qquad (8)$$

At low injection levels, Equation (8) can be reduced to $p_t = \phi \tau_r (\tau_g/\tau_t)$. The influence of the trap states on photoconductivity can thus be understood to result from an increase of effective carrier lifetime by $(\tau_g/\tau_t)$. We insert above equations into (4), calculate the responsivity from (3), and fit the model (solid line in Figure 3d) to the experimental data by adjusting the ratio $(\tau_g/\tau_t) \sim 200$ and the trap density $P_t \sim 5\times10^{10}$ cm$^{-2}$. $P_t$ is similar to that obtained by space–charge limited current measurements[40], confirming the validity of our assumptions. The effective energetic position of the traps can now be estimated from[44] $E_{V,t} = kT \ln[N_V/P_t \times \tau_g/\tau_t] \sim 0.27$ eV (above $E_V$). $k$ is Boltzmann's constant, $N_V \sim g_v m^* kT/(\pi \hbar^2) \sim 8.6 \times 10^{12}$ cm$^{-2}$ is the 2D effective density of states in the valence band, $g_v = 2$ is the valley degeneracy, $m^* \sim 0.4 \times m_0$ is the effective mass[41], and $m_0$ denotes the free electron mass.

The decrease of photocurrent in the transistor OFF-state (symbols in Figure 3e) is attributed to trap states near the conduction band edge (band tail) that become emptied as the Fermi level shifts towards mid-gap. Their influence can be lumped into the effective mobility $\mu_{n,eff}$, which is the mobility of free electrons in the band reduced by the fraction of carriers in the traps[46],

$$\mu_{n,eff} = \frac{\mu_n}{1 + \frac{N_t - n_t}{N_C} \exp\left(\frac{E_{C,t}}{kT}\right)}. \quad (9)$$

The tail is again approximated by a narrow distribution of states with density $N_t \sim 5 \times 10^{10}$ cm$^{-2}$ and with energy $E_{C,t}$ below $E_C$. $N_C \sim 8.6 \times 10^{12}$ cm$^{-2}$ is the effective density of states in the conduction band. The term $N_t - n_t$ describes the density of unoccupied trap states, determined by the Fermi level. In the transistor ON-state, $n_t = N_t$ and $\mu_{n,eff} = \mu_n$, as expected. In the OFF-state ($V_G \ll V_T$), on the contrary, $n_t = 0$ and the effective mobility drops according to Equation (9). By fitting the model to the experimental data (solid lines in Figure 3e), we extract $E_{C,t} \sim 0.2$ eV. We note that this value approximately agrees with binding energies of donor levels in MoS$_2$ determined by electrical transport measurements[45]. The theory reproduces rather well the main features of the experiment, such as the drop of the photocurrent in the OFF-state and the flattening of the gate voltage dependence under strong optical illumination due to trap state filling. The more abrupt transition is due to our assumption of a discrete distribution of traps instead of a continuous one.

Finally, we summarize PC results obtained from 6 different bilayer devices (Samples 1–6). The maximum measured responsivity varies strongly from device to device, and exhibits values between 0.25 and 4.1 A/W (inset in Figure 4). Below saturation and for homogeneous illumination ($A = WL$), it is straightforward to show that

$$R = \eta \frac{\lambda q}{hc} \times \frac{\tau_{eff}}{t_{tr}}, \qquad (10)$$

where $\tau_{eff} = \tau_r(\tau_g/\tau_t)$ is the effective carrier lifetime and $t_{tr} = L^2/(\mu_n V_D)$ is the electron transit time through the device. The latter term in Equation (10) is the number of carriers passing the device per photogenerated electron-hole pair and is referred to as photoconductive gain. If we now plot the photoresponsivity $R$ against the inverse carrier transit time $t_{tr}^{-1}$, a well-defined relation is obtained (see Figure 4). The approximately linear scaling of $R$ with $\mu_n$ and $L^{-2}$ provides further evidence of the validity of our model.

In summary, we have studied the photoconductivity of atomically thin layers of $MoS_2$. We have identified a PV effect, which stems from a shift in transistor threshold voltage due to charge transfer, and a PC effect, that is attributed to hole trapping in $MoS_2$ band tail states.

## ASSOCIATED CONTENT

**Supporting Information**

Sample parameters, discussion of alternative photocurrent generation mechanisms, and photocurrent maps. This material is available free of charge via the Internet at http://pubs.acs.org.

## AUTHOR INFORMATION

**Corresponding Author**

*E-mail: thomas.mueller@tuwien.ac.at. Phone: 43-1-58801-38739.

**Notes**

The authors declare no competing financial interest.

## ACKNOWLEDGEMENTS

We thank E. Bertagnolli and A. Lugstein for providing access to a Raman spectrometer. The research leading to these results received funding from the Austrian Science Fund FWF (START Y-539) and the European Union Seventh Framework Programme (grant agreement no. 604391 Graphene Flagship).

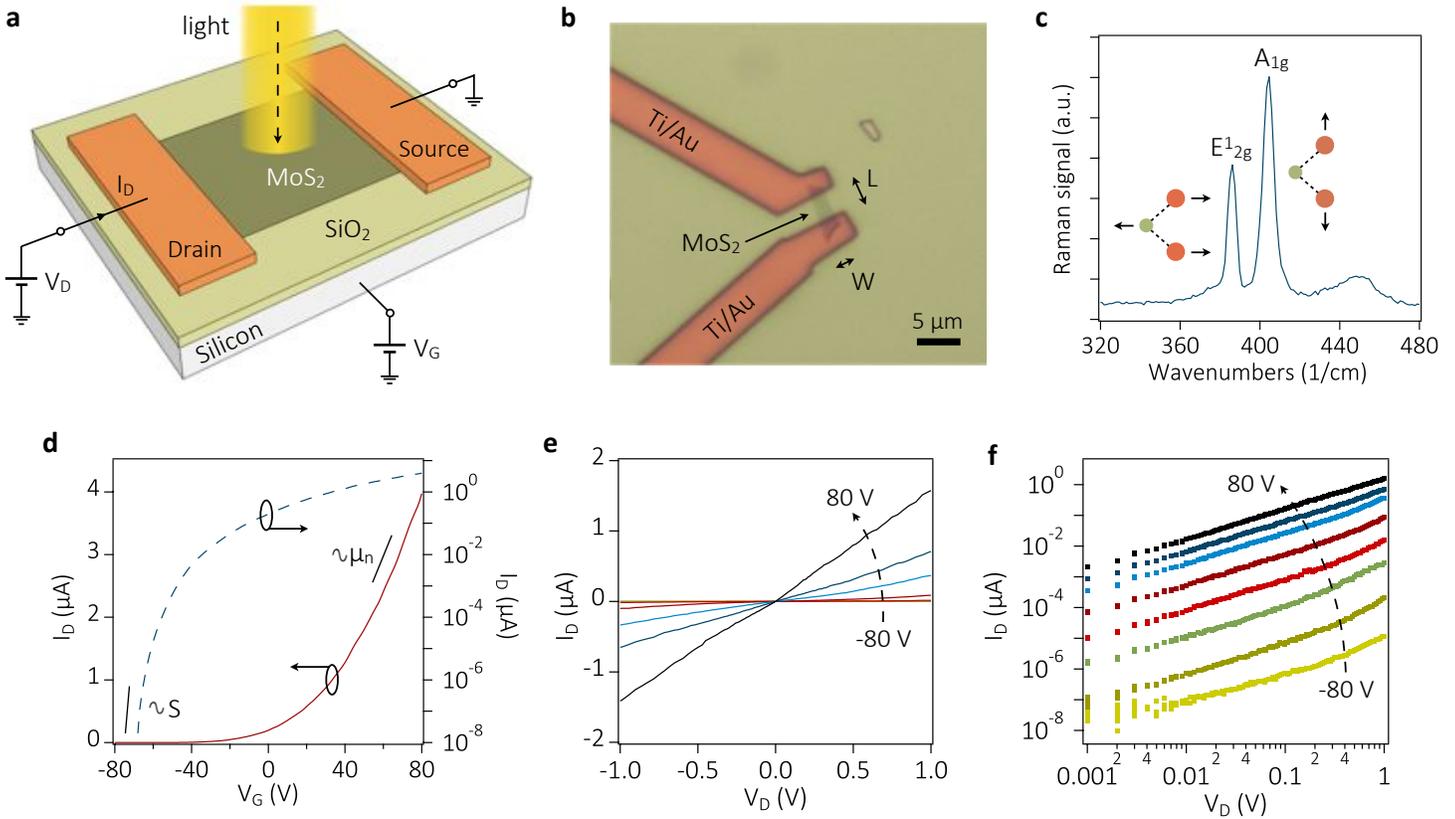

**Figure 1.** (a) Schematic drawing of the experiment. The gate voltage is applied to the silicon substrate. (b) Microscope image of Sample 7 (taken before metal evaporation and lift-off). (c) Raman spectrum of a MoS$_2$ monolayer. The energy difference between the in-plane $E^1_{2g}$ and the out-of-plane $A_{1g}$ Raman modes is ~19 cm$^{-1}$, confirming monolayer thickness. (d) $I_D$–$V_G$ characteristic (Sample 7), plotted on a linear scale (solid red line; left axis) and on a logarithmic scale (dashed blue line, right axis). $V_D$ = 2 V. (e) $I_D$–$V_D$ characteristics (Sample 7) for gate voltages of $V_G$ = -80, -70, -50, -30, 0, 30, 50, and 80 V. (f) Same data as in d, but plotted on a logarithmic scale.

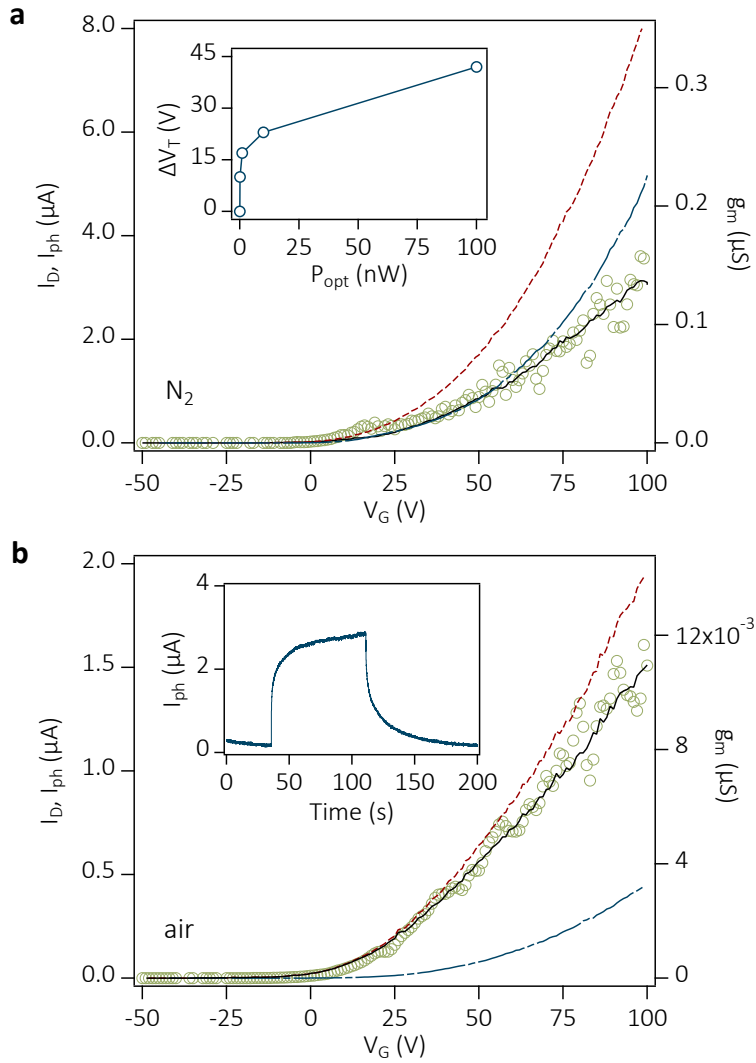

**Figure 2.** (a) $I_D$–$V_G$ measurements performed in the dark (dash-dotted blue line) and under illumination with $P_{opt} \sim 5$ nW (dashed red line). Measurements were performed in $N_2$-atmosphere and with $V_D = 5$ V. The photocurrent is shown as solid black line. The transconductance is plotted in the same graph on the right axis (symbols). Inset: Dependence of $\Delta V_T$ (extracted from $I_D$–$V_G$ measurements) on incident optical power $P_{opt}$. (b) Measurement in air. The meaning of the curves is the same as in b. Inset: Photocurrent time-trace recorded at $V_G = 100$ V.

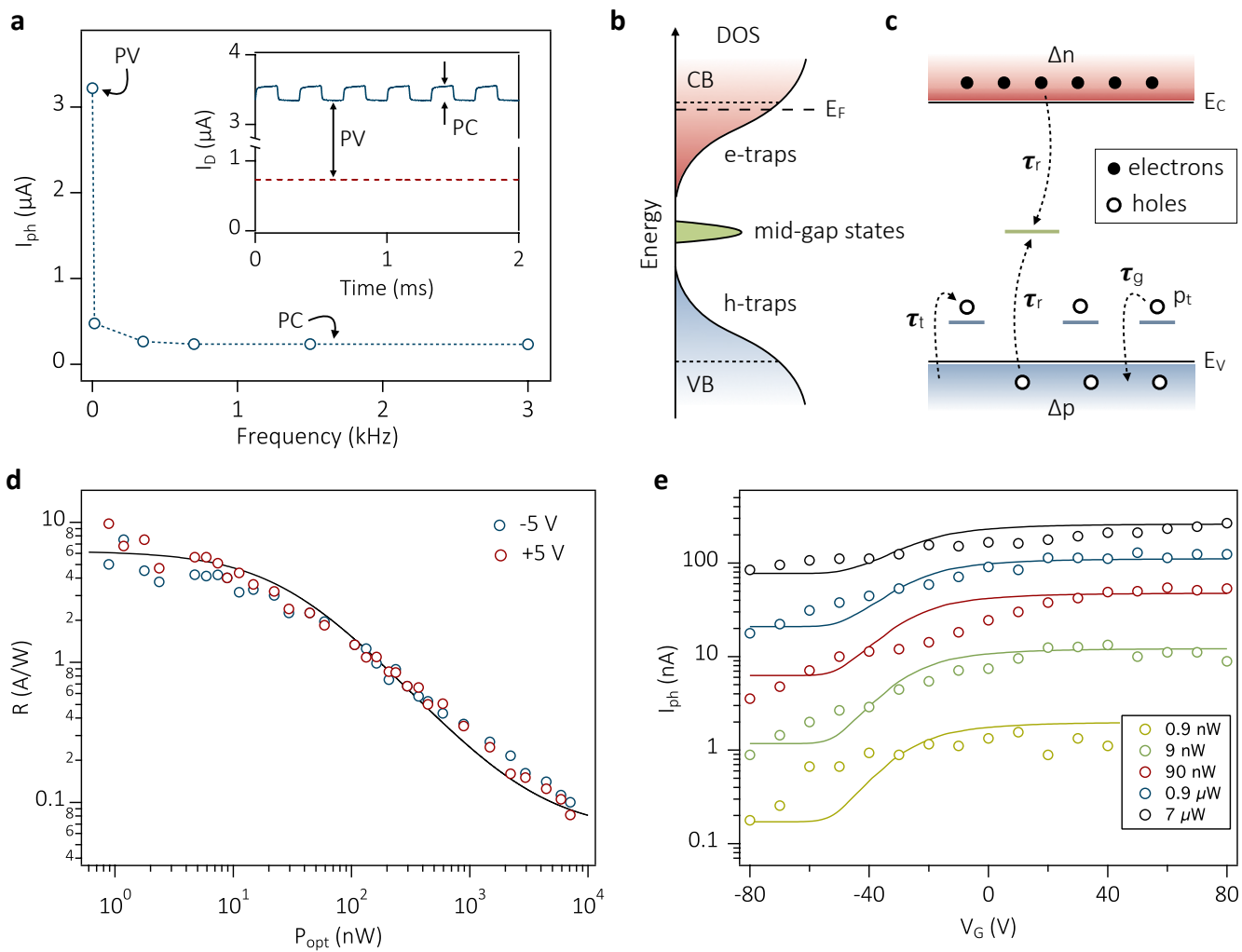

**Figure 3.** (a) Photocurrent amplitude versus modulation frequency. The first data point is the DC measurement (PV effect). The PC effect shows no modulation frequency dependence in the measured range (up to 3 kHz). Inset: $I_D$-time traces taken in the dark (dashed red line) and under illumination (solid blue line). (b) Schematic drawing of the density-of-states (DOS) in atomically thin MoS$_2$. CB, conduction band; VB, valence band. Band tail states exist underneath (above) the conduction (valence) band edge that act as electron (hole) charge traps. The VB-DOS is assumed to be mirror-symmetric to the CB-DOS. We assume the recombination to occur via mid-gap states with an empirical (constant) rate $\tau_r^{-1}$. (c) Simplified energy band diagram, showing the main features of the charge trapping model. The VB tail is approximated by a discrete distribution of hole traps with density $P_t$ (occupation $p_t$) and energy $E_{V,t}$ above the valence band edge $E_V$. $\tau_t^{-1}$ and $\tau_g^{-1}$ are the hole trapping and release rates. $E_F$, Fermi level. (d) Power dependence of photoresponsivity at $V_D$ = -5 V (blue symbols) and $V_D$ = +5 V (red symbols). Solid line: theoretical results. (e) Gate voltage dependence of photocurrent ($V_D$ = 2 V) for optical powers between 0.9 nW and 7 μW. Symbols: measurements; solid lines: theoretical results.

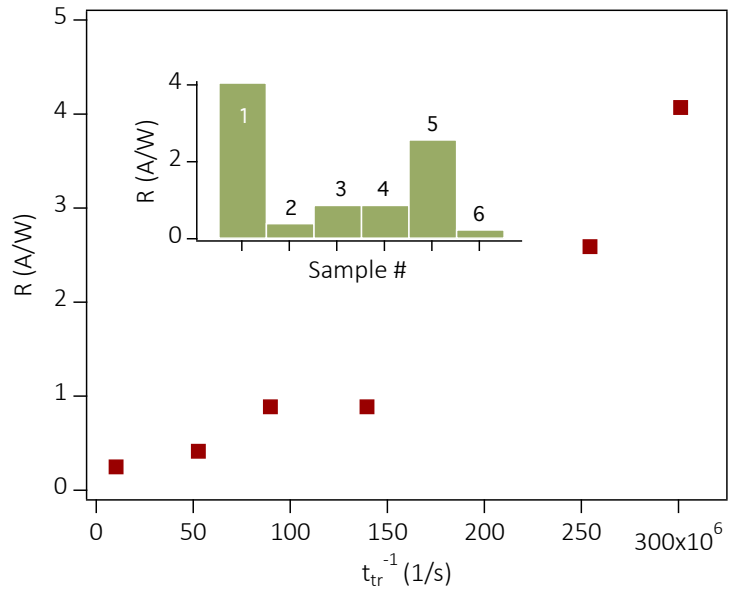

**Figure 4.** Photoresponsivity $R$ versus inverse carrier transit time $t_{tr}^{-1}$. Inset: Photoresponsivities obtained from six different MoS$_2$ bilayer devices.

*Supporting Information for*

# Mechanisms of photoconductivity in atomically thin MoS$_2$

Marco M. Furchi, Dmitry K. Polyushkin, Andreas Pospischil, and Thomas Mueller[*]

*Vienna University of Technology, Institute of Photonics,*
*Gußhausstraße 27-29, 1040 Vienna, Austria*

[*]E-mail: *thomas.mueller@tuwien.ac.at*

| Sample | Type | W (μm) | L (μm) | $\mu_n$ (cm$^2$/Vs) |
|---|---|---|---|---|
| 1 | BL | 5.8 | 1.6 | 1.6 |
| 2 | BL | 1.6 | 3.9 | 1.6 |
| 3 | BL | 3.5 | 2.5 | 1.8 |
| 4 | BL | 3.0 | 3.3 | 2.0 |
| 5 | BL | 6.0 | 2.7 | 3.6 |
| 6 | BL | 5.0 | 7.0 | 0.9 |
| 7 | ML | 3.0 | 1.8 | 7.6 |

**Table S1:** *Device dimensions (channel width W and length L), electron mobility $\mu_n$, and MoS$_2$ layer thickness (ML, monolayer; BL, bilayer) for different samples.*

*Photoelectric (PE) effect.* Assuming 100 % internal quantum efficiency and $\eta = 15$ % optical absorption, we estimate a maximum achievable responsivity of $R_{PE} = \frac{\eta \lambda q}{hc} = 60$ mA/W. At low $P_{opt}$, this value is two orders of magnitude smaller than our experimental observations. The PE effect may become important, though, at high illumination intensities[21] (without the scope of this work).

*Photothermoelectric (PTE) effect.* The photovoltage $V_{PTE}$ at the MoS$_2$/metal junction generated by the photothermoelectric effect can be written as $V_{PTE} = (S_C - S_M)\Delta T$, where $S_C$ ($\ll S_M$) and $S_M$ are the Seebeck coefficients of the contact metal and MoS$_2$, respectively, and $\Delta T$ is the temperature difference at the junction. We calculate the Seebeck coefficient of MoS$_2$ from the electrical transport data using the Mott relation $S_M = -\frac{\pi^2 k^2 T}{3q} \frac{1}{G} \frac{dG}{dV_G} \left(\frac{dV_G}{dE}\right)_{E=E_F}$, where the conductance $G = I_D/V_D$ and its derivative $dG/dV_G$ are taken from the IV-curve in Figure 1d. The term $(dV_G/dE)_{E=E_F}$ is estimated from the relation between Fermi level and gate voltage, $E_F = \frac{\hbar^2 \pi C_{ox}}{2qm^*}(V_G - V_T)$. At $V_G = 80$ V, we obtain $S_M = -0.85$ mV/K, which is similar to results reported previously[23]. The temperature difference $\Delta T/P_{opt} \approx 2.5$ K/mW is taken from Ref. 23 where FEM simulations were performed for a similar device geometry. We then estimate a photoresponsivity of $R_{PTE} \approx S_M(\Delta T/P_{opt})G \approx 0.01$ mA/W, which is several orders of magnitude lower than our experimental results. For lower gate voltages, we calculate even smaller $R_{PTE}$.

*Bolometric (BOL) effect.* The conductivity of atomically thin $MoS_2$ varies strongly with temperature due to thermal activation of trapped electrons. Optical heating of $MoS_2$ may thus lead to a bolometric response. We estimate the bolometric effect from[*] $G = G_0 \exp(-E_a/kT)$, where $E_a \approx 0.1$ eV is a rough estimate of the activation energy. For simplicity, $G_0$ is assumed to be temperature-independent. We obtain $R_{BOL} = \frac{E_a}{kT^2}(\Delta T/P_{opt})I_D \approx 0.1$ mA/W (at $I_D = 4$ µA), with $\Delta T/P_{opt}$ taken from Ref. 23. $R_{BOL}$ is again too small to explain our experimental results.

*Schottky barrier lowering.* Photoconductive gain can be caused by carrier trapping in the vicinity of the $MoS_2$/metal junction, resulting in a lowering of the Schottky barrier under illumination[32]. We exclude this effect, because the current flow in our devices is not controlled by the contacts (see Figures 1e and f).

*Note S1:* Discussion of alternative photocurrent generation mechanisms.

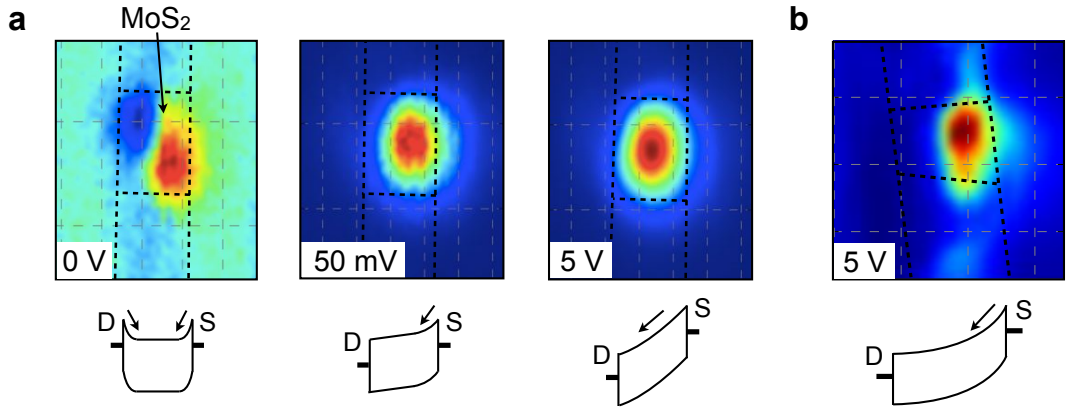

*Figure S1:* (a) Photocurrent maps (amplitude normalized) recorded for different drain-source voltages (Sample 3). For zero bias (left image), very weak PE/PTE signals are obtained at the contacts. The PC response ($V_D \neq 0$) extends over a rather large area across the $MoS_2$ flake. (b) In some devices (especially those with longer channel lengths; e.g. Sample 6) the PC maximum occurs closer to one of the contacts due to the non-uniform voltage drop along the channel.

---

[*] Radisavljevic, B; Kis, A. *Nature Mater.* **2013**, 12, 815–820.